\documentclass[conference]{IEEEtran}
\IEEEoverridecommandlockouts

\usepackage{cite}
\usepackage{amsmath,amssymb,amsfonts}
\usepackage{algorithmic}
\usepackage{graphicx}
\usepackage{xfp}
\usepackage{textcomp}
\usepackage{xcolor}
\usepackage{xfrac}
\usepackage{enumitem}
\usepackage{caption}
\usepackage{subcaption}
\usepackage{balance}
\usepackage{makecell}
\usepackage{comment}
\usepackage{multirow}
\usepackage{tabularx}
\usepackage{pgfplots}
\pgfplotsset{compat=newest}

\usepackage[acronym]{glossaries}
\newglossary[symlog]{symbol}{symi}{symo}{Symbols}
\makeglossaries
\glstoctrue 
\newcommand{\newacronymf}[3]{\newglossaryentry{#1}{name={#2},description={#3},first={#2}}}

\newacronym{3gpp}{3GPP}{Third Generation Partnership Project}
\newacronym{4g}{4G}{fourth generation of mobile phone system specifications}
\newacronym{5g}{5G}{fifth generation of mobile phone system specifications}
\newacronym{1dlstm}{1D-LSTM}{One-dimensional Long Short-Term Memory}
\newacronym{2dlstm}{2D-LSTM}{Two-dimensional Long Short-Term Memory}
\newacronym{1dcnn}{1D-CNN}{One-dimensional Convolutional Neural Network}
\newacronym{2dcnn}{2D-CNN}{Two-dimensional Convolutional Neural Network}

\newacronym{ann}{ANN}{Artificial Neural Network}
\newacronym{anns}{ANNs}{Artificial Neural Networks}
\newacronym{acq}{ACQ}{acquisition}
\newacronym{app}{APP}{a posteriori probability}
\newacronym{ask}{ASK}{amplitude-shift keying}
\newacronym{awgn}{AWGN}{Additive White Gaussian Noise}
\newacronym{arq}{ARQ}{automatic repeat request}
\newacronym{acs}{ACS}{add-compare-select}
\newacronym{asic}{ASIC}{Application-Specific Integrated Circuit}
\newacronym{arp}{ARP}{almost regular permuatation}
\newacronym{asip}{ASIP}{application specific instruction set processor}
\newacronym{amba}{AMBA}{Advanced Microcontroller Bus Architecture}
\newacronym{ahb}{AHB}{advanced high-performance bus}
\newacronym{aacs}{AACS}{a priori aware constant sphere}
\newacronym{ant}{ANT}{algorithmic noise-tolerance}
\newacronym{ai}{AI}{Artificial Intelligence}
\newacronym{af}{AF}{Atrial Fibrillation}
\newacronym{aecg}{AECG}{Ambulatory Electrocardiographic}

\newacronym{bhl}{BHL}{Backward Hidden Layer}
\newacronym{bcjr}{BCJR}{Bahl, Cocke, Jelinek, and Raviv}
\newacronym{bpsk}{BPSK}{Binary Phase Shift Keying}
\newacronym{ber}{BER}{bit error rate}
\newacronym{bp}{BP}{Belief Propagation}
\newacronym{bw}{BW}{Bit Width}
\newacronym{bicm}{BICM}{bit interleaved coded modulation}
\newacronym{blast}{BLAST}{Bell Labs layered space time}
\newacronym{bc}{BC}{best case, 1.3V, -40$^\circ$C}
\newacronym{bilstm}{Bi-LSTM}{Bidirectional LSTM}
\newacronym{bmbf}{BMBF}{Federal Ministry of Education and Research}
\newacronym{bnn}{BNN}{Binarized Neural Network}
\newacronym{bram}{BRAM}{Block RAM}

\newacronym{ctc}{CTC}{Connectionist Temporal Classification}
\newacronym{cn}{CN}{Check Node}
\newacronym{cr}{CR}{Code Rate}
\newacronym{cfu}{CFU}{Check Node Functional Unit}
\newacronym{crc}{CRC}{Cyclic Redundancy Check}
\newacronymf{cdma2000}{CDMA2000}{name of a communication standard}
\newacronym{cc}{CC}{convolutional code}
\newacronym{cnb}{CNB}{check node block}
\newacronym{cnp}{CNP}{check node processor}
\newacronym{cdma}{CDMA}{code division multiple access}
\newacronym{cmos}{CMOS}{complementary metal oxide semiconductor}
\newacronym{cpu}{CPU}{Central Processing Unit}
\newacronym{cnn}{CNN}{Convolutional Neural Network}
\newacronym{cer}{CER}{Character Error Rate}
\newacronym{ccl}{CCL}{Connected Component Labeling}
\newacronym{clb}{CLB}{Configurable Logic Block}

\newacronym{darts}{DARTS}{Differentiable ARchiTecture Search}
\newacronym{dart}{DART}{Distributed Analytics Runtime}
\newacronym{dma}{DMA}{Direct Memory Access}
\newacronym{dnnu}{DNNU}{Deep Neural Network Unit}
\newacronym{db}{dB}{decibel}
\newacronym{dp}{DP}{dual ported (memory)}
\newacronym{dram}{DRAM}{Dynamic Random Access Memory}
\newacronym{dc}{$d_c$}{CN degree}
\newacronym{dv}{$d_v$}{VN degree}
\newacronym{dvs}{DVS}{dynamic voltage scaling}
\newacronym{dvb}{DVB}{digital video broadcast}
\newacronym{dl}{DL}{Deep Learning}
\newacronym{dnn}{DNN}{Deep Neural Network}
\newacronymf{dvbrcs}{DVB-RCS}{digital video broadcasting - return channel over satellite}
\newacronymf{dvbrct}{DVB-RCT}{digital video broadcasting - return channel terrestrial}
\newacronymf{dvbc2}{DVB-C2}{digital video broadcasting - cable - second generation \cite{eudigital}}
\newacronymf{dvbs2}{DVB-S2}{digital video broadcasting - satellite - second generation \cite{eudigital}}
\newacronymf{dvbs2x}{DVB-S2X}{digital video broadcasting - satellite - extension of the second generation \cite{eudigital}}
\newacronymf{dvbt2}{DVB-T2}{digital video broadcasting - terrestrial - second generation \cite{eudigital}}
\newacronym{dvfs}{DVFS}{dynamic voltage and frequency scaling}
\newacronym{drp}{DRP}{dithered relative prime}
\newacronym{dfg}{DFG}{Deutsche Forschungsgesellschaft}
\newacronym{dibco}{DIBCO}{Document Image Binarization Competition}
\newacronym{dsp}{DSP}{Digital Signal Processing}
\newacronym{dfki}{DFKI}{German Research Centre for Artificial Intelligence}
\newacronym{dpi}{DPI}{Dots Per Inch}
\newacronym{dnndk}{DNNDK}{Deep Neural Network Development Kit}
\newacronym{dpu}{DPU}{Deep Learning Processor Unit}

\newacronym{esf}{ESF}{extrinsic scaling factor}
\newacronym{edc}{EDC}{error detection and correction}
\newacronym{edge}{EDGE}{name of a communications standard}
\newacronym{eds}{EDS}{error detection sequential}
\newacronym{ecc}{ECC}{error correction code}
\newacronym{exit}{EXIT}{extrinsic information transfer}
\newacronym{ems}{EMS}{Extended Min-Sum}
\newacronym{ecn}{ECN}{Elementary Check Node}
\newacronym{ecg}{ECG}{Electrocardiogram}
\newacronym{ebr}{EBR}{Embedded Block RAM}

\newacronym{fhl}{FHL}{Forward Hidden Layer}
\newacronym{fsm}{FSM}{finite state machine}
\newacronym{fec}{FEC}{forward error correction}
\newacronym{fer}{FER}{Frame Error Rate}
\newacronym{fifo}{FIFO}{First In First Out memory}
\newacronym{fpga}{FPGA}{Field-Programmable Gate Array}
\newacronym{fsd}{FSD}{Fixed complexity sphere decoder}
\newacronym{fwbw}{FWBW}{Forward-Backward}
\newacronym{fcn}{FCN}{Fully Convolutional Network}
\newacronym{fpn}{FPN}{Feature Pyramid Network}
\newacronym{ff}{FFs}{Flip-Flops}
\newacronym{fps}{FPS}{Frames Per Second}
\newacronym{fcnn}{FCNN}{Fully Connected Neural Network}
\newacronym{fc}{FC}{Fully Connected}

\newacronym{gsm}{GSM}{Global System for Mobile Communication; communication standard}
\newacronym{gpp}{GPP}{general-purpose processor}
\newacronym{gprs}{GPRS}{name of a communications standard}
\newacronym{gps}{GPS}{global positioning system}
\newacronym{geq}{GE}{gate equivalent}
\newacronym{gfq}{GF$(q)$}{Galois Field}
\newacronym{gpu}{GPU}{Graphics Processing Unit}
\newacronym{gpi}{GPI}{Global address space Programming Interface}
\newacronym{gops}{GOp/s}{Giga Operations per second}
\newacronym{gap}{GAP}{Global Average Pooling}

\newacronym{hspa}{HSPA}{high speed packet access}
\newacronym{harq}{HARQ}{hybrid automatic repeat request}
\newacronym{hw}{HW}{hardware}
\newacronym{hls}{HLS}{High-level Synthesis}
\newacronym{half}{HALF}{\underline{H}olistic \underline{A}uto machine \underline{L}earning for \underline{F}PGAs}
\newacronym{hdl}{HDL}{Hardware Description Language}

\newacronym{ic}{IC}{integrated circuit}
\newacronym{io}{I/O}{input / output}
\newacronym{ip}{IP}{Intellectual Property}
\newacronym{ier}{IER}{injected error rate}
\newacronym{itrs}{ITRS}{International Technology Roadmap for Semiconductors}
\newacronym{iot}{IoT}{Internet of Things}
\newacronym{ilsvrc}{ILSVRC}{ImageNet Large Scale Visual Recognition Challenge}
\newacronym{iou}{IoU}{Intersection over Union}

\newacronym{kde}{KDE}{Kernel Density Estimator}
\newacronym{kd}{KD}{Knowledge Distillation}

\newacronym{lstm}{LSTM}{Long Short-Term Memory}
\newacronym{llr}{LLR}{Logarithmic Likelihood Ratio}
\newacronym{ldr}{LDR}{Logarithmic Density Ratio}
\newacronym{lte}{LTE}{long term evolution of the 3GPP standard}
\newacronym{ldpc}{LDPC}{Low-Density Parity-Check}
\newacronym{logmap}{Log-MAP}{MAP algorithm in the logarithmic domain}
\newacronym{lan}{LAN}{local area network }
\newacronym{lsb}{LSB}{least significant bit}
\newacronym{lfsr}{LFSR}{linear feedback shift register}
\newacronym{lord}{LORD}{layered orthogonal lattice detector}
\newacronym{lsd}{LSD}{List sphere decoder}
\newacronym{ldmc}{LDMC}{low-density MIMO check}
\newacronym{lut}{LUTs}{LookUp Tables}

\newacronym{mac}{MAC}{Multiply-Accumulate}
\newacronym{map}{MAP}{maximum a posteriori probability}
\newacronym{mlm}{Max-Log-MAP}{MAP algorithm in the logarithmic domain with simplified computation}
\newacronym{mimo}{MIMO}{multiple input multiple output}
\newacronym{mtbf}{MTBF}{mean time between failure}
\newacronym{msb}{MSB}{most significant bit}
\newglossaryentry{mpsoc}{name=MPSoC, description=multi-processor system-on-chip, first=multi-processor system-on-chip (MPSoC), firstplural=multi-processor systems-on-chip (MPSoC)}
\newacronym{mops}{MOPS}{million operations per second}
\newacronym{mmse}{MMSE}{minimum mean square error}
\newacronym{mcmc}{MCMC}{Markov Chain Monte Carlo}
\newacronym{mll}{ML}{maximum likelihood}
\newacronym{ml}{ML}{Machine Learning}
\newacronym{mdlstm}{MD-LSTM}{Multidimensional Long Short-Term Memory}
\newacronym{mlp}{MLP}{Multilayer Perceptron}
\newacronym{mdpi}{MDPI}{Multidisciplinary Digital Publishing Institute}

\newacronym{nns}{NNs}{Neural Networks}
\newacronym{nn}{NN}{Neural Network}
\newacronym{nbldpc}{NB-LDPC}{Non-Binary Low-Density Parity-Check}
\newacronym{nii}{NII}{next iteration initialization}
\newacronym{nsc}{NSC}{non-systematic and non-recursive convolutional}
\newacronym{noc}{NoC}{network on chip}
\newacronym{nom}{NOM}{nominal case, 1.2V, 25$^\circ$C}
\newacronym{nas}{NAS}{neural architecture search}
\newacronym{nlp}{NLP}{Natural Language Processing}
\newacronym{npu}{NPU}{Neural Processing Unit}

\newacronym{ocr}{OCR}{Optical Character Recognition}
\newacronym{otx}{OTX}{outer receiver}
\newacronym{ofdm}{OFDM}{orthogonal frequency division multiplexing}

\newacronym{pbb}{PBB}{Percentile-Based Binarization}
\newacronym{pl}{PL}{Programmable Logic}
\newacronym{ps}{PS}{Processing System}
\newacronym{psk}{PSK}{phase-shift keying}
\newacronym{psmap}{PSMAP}{parallel SMAP}
\newacronym{pr}{P\&R}{place and route}
\newacronym{ped}{PED}{partial Euclidean distance}
\newacronym{pic}{PIC}{parallel interference cancellation}
\newacronym{pn}{PN}{Permutation Node}
\newacronym{pim}{PIM}{Processing-In-Memory}
\newacronym{pcb}{PCB}{Printed Circuit Board}
\newacronym{pp}{pp}{percentage point}
\newacronym{ptq}{PTQ}{Post-Training Quantization}

\newacronym{qam}{QAM}{quadrature amplitude modulation}
\newacronym{qat}{QAT}{quantization aware training}
\newacronym{qbn}{QBN}{quantized batch norm}
\newacronym{qpsk}{QPSK}{quadrature phase shift keying}
\newacronym{qpp}{QPP}{quadratic permutation polynomial}
\newacronym{qos}{QoS}{Quality of Service}

\newacronym{rnn}{RNN}{Recurrent Neural Network}
\newacronym{rsc}{RSC}{recursive systematic convolutional}
\newglossaryentry{ram}{name=RAM,description=random access memory,first=random access memory (RAM),firstplural=random access memories (RAMs)}
\newglossaryentry{rom}{name=ROM,description=read only memory,first=read only memory (ROM),firstplural=read only memories (ROMs)}
\newacronym{rms}{RMS}{recognition, mining, and synthesis}
\newacronym{rs}{RS}{reed-solomon code}
\newacronym{rtl}{RTL}{register transfer level}
\newacronym{rts}{RTS}{repeated tree search}
\newacronym{rap}{RAP}{Resilience Articulation Point}
\newacronym{relu}{ReLU}{Rectified Linear Unit}
\newacronym{rl}{RL}{Reinforcement Learning}

\newacronym{snr}{SNR}{Signal-to-Noise Ratio}
\newglossaryentry{sram}{name=SRAM,description=static random access memory,first=static random access memory (SRAM),firstplural=static random access memories (SRAMs)}
\newacronym{smap}{SMAP}{serial MAP}
\newacronym{sdr}{SDR}{software-defined radio}
\newacronym{sdram}{SDRAM}{synchronous dynamic random access memory}
\newacronym{set}{SET}{single-event transient}
\newacronym{seu}{SEU}{single-event upset}
\newacronym{siso}{SISO}{soft-input soft-output}
\newacronym{siso2}{SISO}{single-input single-output}
\newacronym{soc}{SoC}{Systems-on-Chip}
\newacronym{sova}{SOVA}{soft-output Viterbi algorithm}
\newacronym{sw}{SW}{software}
\newacronym{sm}{SM}{spatial multiplexing}
\newacronym{stbc}{STBC}{space-time block code}
\newacronym{sttc}{STTC}{space-time trellis code}
\newacronym{sic}{SIC}{successive interference cancellation}
\newacronym{sts}{STS}{single tree search}
\newacronym{spp}{SPP}{Schwerpunktprogramm}
\newacronym{se}{SE}{Schnorr-Euchner}
\newacronym{synb}{SYN}{Syndrome-Based}
\newacronym{sgd}{SGD}{Stochastic Gradient Descent}
\newacronym{sdk}{SDK}{Software Development Kit}

\newacronym{tmr}{TMR}{triple modular redundancy}
\newacronym{tc}{TC}{turbo code}
\newacronym{ts}{TS}{Tuple search}
\newacronym{tpu}{TPU}{Tensor Processing Unit}

\newacronym{umts}{UMTS}{universal mobile telecommunications system}
\newacronym{uwb}{UWB}{ultra-wide band}
\newacronym{usb}{USB}{universal serial bus}
\newacronym{umic}{UMIC}{ultra high speed mobile information and communication}
\newacronym{uav}{UAV}{Unmanned Aerial Vehicle}

\newacronym{vn}{VN}{Variable Node}
\newacronym{vfu}{VFU}{variable node functional unit}
\newacronym{vhdl}{VHDL}{very high speed integrated circuits hardware description language}
\newacronym{vliw}{VLIW}{very large instruction word}
\newacronym{vnb}{VNB}{variable node block (in \gls{LDPC} decoder architectures)}
\newacronym{vnp}{VNP}{variable node processors (in \gls{LDPC} decoder architectures)}
\newacronym{vblast}{V-BLAST}{Vertical Bell Labs layered space time}

\newacronym{wimax}{WiMAX}{worldwide interoperability for microwave access \cite{ieworldwide}}
\newacronym{wimedia}{WiMedia}{name of an alliance for standardization}
\newacronym{wifi}{WiFi}{trademark of Wi-Fi alliance}
\newacronym{wc}{WC}{worst case, 1.1V, 125$^\circ$C}

\newglossaryentry{xmap}{name=XMAP,description={name of a pipelined MAP decoder},first=XMAP,firstplural=XMAPs}
\newacronym{xor}{XOR}{exclusive OR function}

\newacronym{zf}{ZF}{zero-forcing}

\usepackage{booktabs}
\usepackage{threeparttable}
\usepackage{siunitx}

\usepackage{cleveref}

\crefname{figure}{Fig.}{Figs.}
\Crefname{figure}{Fig.}{Figs.}

\crefname{table}{Table}{Tables}
\Crefname{table}{Table}{Tables}

\crefname{section}{Section}{Sections}
\Crefname{section}{Section}{Sections}
\def\BibTeX{{\rm B\kern-.05em{\sc i\kern-.025em b}\kern-.08em
    T\kern-.1667em\lower.7ex\hbox{E}\kern-.125emX}}

\begin{document}

\title{A Case Study on Energy-Efficient Edge AI \\Crack Segmentation\\

\thanks{The author(s) declare that financial support was received for the research, authorship, and/or publication of this article. This work was supported by the European Union's Horizon Europe research and innovation programme (HORIZON-CL4-2021-HUMAN-01) in the project SustainML (101070408).

* These authors contributed equally to this work.}
}
\author{
\IEEEauthorblockN{
Matthias Tschöpe\textsuperscript{*,1},
Mohamed Moursi\textsuperscript{*,2},
Vladimir Rybalkin\textsuperscript{*,2},
Bo Zhou\textsuperscript{1},
Norbert Wehn\textsuperscript{2},
Paul Lukowicz\textsuperscript{1}
}
\IEEEauthorblockA{\textsuperscript{1}DFKI, Kaiserslautern, Germany\\
\{Matthias.Tschoepe, Bo.Zhou, Paul.Lukowicz\}@dfki.de}
\IEEEauthorblockA{\textsuperscript{2}RPTU, Kaiserslautern, Germany\\
\{Mmoursi, Vladimir.Rybalkin, Norbert.Wehn\}@rptu.de}
}

\maketitle
\newcommand\copyrighttext{%
  \footnotesize \textcopyright \the\year{} IEEE. Personal use of this material is permitted. Permission from IEEE must be obtained for all other uses, including reprinting/republishing this material for advertising or promotional purposes, collecting new collected works for resale or redistribution to servers or lists, or reuse of any copyrighted component of this work in other works.}

\newcommand\copyrightnotice{%
\begin{tikzpicture}[remember picture,overlay]
\node[anchor=south,yshift=10pt] at (current page.south) {\fbox{\parbox{\dimexpr0.75\textwidth-\fboxsep-\fboxrule\relax}{\copyrighttext}}};
\end{tikzpicture}%
}

\copyrightnotice
\begin{abstract}

\glsresetall
Crack segmentation on edge devices can support continuous infrastructure monitoring and maintenance and thereby help to preserve public safety. Furthermore, autonomous infrastructure monitoring by using \glspl{uav} can reduce inspection risks, as human operators no longer need to enter hazardous areas. Edge processing reduces the cost of inspection by eliminating the need for high resolution image storage for offline processing and mitigates the security risks and bandwidth requirements of streaming to cloud servers. Edge inference is difficult due to the limited memory and computational capabilities of edge devices, which can affect both accuracy and latency. Furthermore, battery-powered devices are subject to strict power and energy constraints. Together, these limitations impose restrictions on the model size and computational complexity that can be deployed close to the sensor. Recent Transformer-based models reach high accuracy on semantic segmentation, but they are typically large and computationally intensive, which makes them hard to deploy on edge devices. We therefore use knowledge distillation to improve the segmentation results of smaller base models. We then use \gls{ptq} to compress the models further. Additionally, we consider the deployment of these models across multiple edge platforms. To maximize energy efficiency, we design and implement a custom hardware architecture for the models on an FPGA. Our results show that \gls{kd} improves all tested U-Net variants. Among the evaluated platforms, the selected FPGA implementation achieves $398$ \gls{fps} at $204.99$ Frames/J while maintaining a mean \gls{iou} of $69.42\%$. In addition, our best model reaches $71.92\%$ mean \gls{iou}, which is $8.82$ \glspl{pp} higher than the previously reported result on the CrackVision12K dataset.

\end{abstract}


\section{Introduction}
Inspection and maintenance of civil infrastructure, such as bridges, tunnels, and high-rise buildings, are essential to ensure long-term safety and reliability. Traditional inspection is mostly manual: human experts have to access dangerous locations to look for structural damage. With the emergence of \glspl{uav}, automated large-scale inspection of concrete and asphalt surfaces has become feasible, offering both safety and efficiency advantages. \gls{uav}-based inspection systems equipped with onboard computer vision can identify cracks and other forms of surface degradation without exposing human inspectors to hazardous environments.

The main challenge for onboard crack segmentation is the need to operate on resource-constrained edge devices with limited power, memory, and computational capacity, while still enabling real-time processing. Transformer-based models reach high segmentation accuracy, but their compute and memory demands make them difficult to run on edge devices. Lightweight architectures like U-Net are therefore the more practical choice. However, U-Net variants differ a lot in parameter count and compute cost, and still need to be optimized for edge use.
Identifying models that balance segmentation accuracy and efficiency is therefore critical for real-world edge crack segmentation.

Efficient deployment of crack segmentation models on edge devices is governed by strict constraints on latency, power consumption, and energy efficiency, particularly under small batch sizes dictated by the number of input sensors (e.g., cameras). In this work, we look at the implementation on different edge platforms, including CPUs, GPUs, and \glspl{npu}, which all have different tradeoffs. GPUs provide high throughput, especially for large batch sizes, but typically incur higher power consumption and are less efficient in small-batch regimes. CPUs offer greater flexibility and broader operator support, yet often fall short in terms of parallel efficiency. \glspl{npu}, while promising superior energy efficiency, are constrained by limited support for operators and model architectures, complicating deployment. These disparities make platform selection, numerical precision, and runtime configuration non-trivial design choices. Collectively, these limitations highlight the need for custom \gls{fpga}-based architectures to achieve superior energy efficiency and mitigate platform-specific constraints.

Our contributions are as follows:

\begin{itemize}
    \item We explore the design space over model scaling, numerical precision, and hardware platform, and we compare segmentation accuracy, latency, power, and energy efficiency for crack segmentation on four edge platforms (CPU, GPU, TPU, and FPGA).
    
    \item We design and implement U-Net variants spanning three orders of magnitude in parameter count to study the trade-off between segmentation performance and efficiency. We use \gls{kd} from a Transformer model (teacher) to improve the smaller U-Net model (student). Under the train/val/test split released in \cite{goo2025hybrid} for CrackVision12K, our most accurate model reaches 8.82\glspl{pp} higher mean \gls{iou} than their Hybrid-Segmentor model, while using $30.64\times$ fewer parameters.
    
    \item To further improve energy efficiency and throughput for single-batch processing, we develop a custom hardware architecture and implement it on an FPGA.
    
    \item Our Pareto-optimal FPGA implementation reaches the highest energy efficiency (204.99 Frames/J) and throughput (398 \gls{fps}) of all tested platforms, while also improving the mean IoU by 2.79\glspl{pp} over the baseline U-Net without \gls{kd}.
\end{itemize}

\section{Related Work}
\subsection{Crack Segmentation Approaches}
Early methods of crack segmentation are often based on graph-based approaches, such as shortest path algorithms \cite{gunkel2012micro} or minimum spanning trees \cite{zou2012cracktree}. These methods are efficient and explainable, but since the introduction of U-Net \cite{ronneberger2015u}, deep learning models recognize cracks much better. Due to its efficient encoder-decoder structure with skip connections, U-Net has become particularly popular for crack segmentation, especially for real-time applications. As a result, there are now many deep learning approaches that build on this and propose new, improved models or training procedures \cite{choi2019sddnet, liu2023crackformer, han2021crackw, lin2023deepcrackat, wang2022automatic, lau2020automated, egodawela2024surface}. U-Net++ was introduced in 2019 \cite{zhou2019unet++}, and in 2023, an improved U-Net incorporating an attention mechanism was proposed \cite{al2023improved}. However, most of these variants require increased computational resources and are therefore not suitable for edge platforms. One of the most advanced real-time capable models is UNext \cite{chang2024unext}, but its reliance on attention mechanisms introduces architectural complexity that currently hinders edge deployments. More recent work explores Mamba- and state-space-model-based crack segmentation \cite{liu2025scsegamba}, which reduces the parameter count further. In our work, we instead start from a standard U-Net and recover the accuracy of compact variants via knowledge distillation, an approach that has also been shown to be effective for crack segmentation in \cite{diasdacosta2024robust}.

\subsection{Semantic Segmentation on Edge Platforms}
In recent years, edge platforms have gained significant traction for crack segmentation. Some authors propose specifically designed segmentation models for edge platforms. For example, \cite{chen2023lightweight} presents MCFF-L Net - an architecture with 1.18 million parameters - combined with knowledge distillation, which achieves good segmentation results and is able to run in real-time on a Jetson Xavier NX. Similarly, \cite{zhang2025device} considered crack segmentation on devices for structural health monitoring. A related approach by \cite{zhang2025crack} uses a special \gls{kd} approach for measurement tasks on edge systems.

Other works focus on simplifying U-Net for microcontroller-class hardware, like \cite{falaschetti2022lightweight}, which analyzed small CNN-based crack segmentation models on low-power microcontrollers. In contrast, \cite{falaschetti2023u} showed that U-Net models can be compressed to meet the requirements of very resource-constrained devices such as the STM32 platform.

\subsection{Semantic Segmentation on FPGA}
Recently, \glspl{fpga} have become a popular choice for implementing semantic segmentation algorithms under real-time and limited power budget constraints. In \cite{ghielmetti2022real}, a compressed ENet is deployed entirely on-chip using HLS4ML, while \cite{jia2021design} and \cite{shen2024fpga} combine pruning, quantization, and Vitis-AI flow to run encoder-decoder networks in real time on \gls{fpga}. Other approaches redesign segmentation blocks specifically for hardware efficiency, such as \cite{chen2024fpga}. Another approach to shrinking the model is about model compression and low-bit quantization. In \cite{mori2022accelerating}, DeepLabV3+ is pruned using a hardware-aware genetic algorithm, in \cite{Modiboyina} a U-Net is quantized down to 4-bit and implemented using a shift based accelerator, while in \cite{miyama2021fpga} a U-Net is quantized to 3-bit precision and implemented on a dedicated on-chip pipeline. Similar OpenCL- or HLS-based encoder-decoder accelerators further show the impact of quantization and optimized data flow on performance \cite{yu2019optimizing}. However, in the context of crack segmentation, \gls{fpga}-based solutions are less common. One of the few works in this direction is \cite{chisholm2019fpga} , which implements a particle-filter-based crack detector directly on an FPGA for real-time inspection. Although they use a totally different approach than we do, their approach already shows the relevance of FPGAs for building inspections.

\subsection{\gls{fpga}-based U-Net Implementations}

Existing work on U-Net deployment on \glspl{fpga} can be broadly categorized into two architectural paradigms: reconfigurable (engine-based) architectures and fully streaming (dataflow) architectures.

Reconfigurable architectures instantiate a limited set of generic compute engines in the \gls{pl}, which are time-multiplexed across network layers. Each layer is executed by reprogramming the compute engines with the corresponding parameters and supplying the required input feature maps. This approach, explored in \cite{realtime_seg}, \cite{fpga_ultrasound}, and \cite{hsi_seg}, provides high flexibility and enables the deployment of networks whose size exceeds on-chip resource capacity. However, it incurs frequent off-chip memory transfers for intermediate feature maps and parameters, leading to increased memory bandwidth requirements and higher energy consumption.

In contrast, streaming architectures map the entire network onto the \gls{pl} as a sequence of dedicated processing modules, enabling pipelined execution with minimal intermediate storage. This dataflow-oriented design significantly reduces memory traffic and improves energy efficiency by exploiting on-chip data reuse. However, the approach is constrained by the available on-chip resources, particularly block RAM and logic, which must accommodate both model parameters and intermediate activations.

U-Net architectures pose additional challenges for streaming implementations due to their encoder–decoder structure with multiple skip connections that require buffering high-resolution feature maps. These skip connections substantially increase on-chip memory demand and can become the primary bottleneck in fully unrolled designs based on widely used FINN \cite{umuroglu2017finn} and HLS4ML \cite{duarte2018fast} hardware libraries. To address this limitation, we propose offloading skip connection feature maps to external memory. This adds extra memory accesses, but the access pattern is mostly sequential and predictable, so we can use the memory bandwidth efficiently. We therefore expect only a small energy overhead, while we can free up enough on-chip memory to fit larger U-Net variants.

\section{Methodology}

\subsection{U-Net Channel Scaling}
We use the original U-Net model \cite{ronneberger2015u} as a basis and generate four smaller variants by reducing the number of feature maps. For this purpose, we introduce a parameter $c \in \{2,4,8,16,32\}$ which controls the number of all feature maps in the network while preserving the original encoder–decoder architecture. Setting $c=32$ reproduces the original U-Net architecture. This parametrization allows us to analyze how the number of feature maps influences the segmentation results as well as the runtime and power consumption.

Each encoder stage consists of two $3\times3$ convolutions with padding $1$, each followed by Batch Normalization and a ReLU activation. Downsampling is done by using a $2\times2$ max-pooling. The decoder mirrors this structure, using either nearest upsampling (only for Edge TPU) or learnable $2\times2$ transposed convolutions. After upsampling, the decoder concatenates the feature maps with the corresponding encoder activations via skip connections and also applies two $3\times3$ convolutions with padding $1$, where each is followed by Batch Normalization and a ReLU activation. Finally, the network uses a $1\times1$ convolution to map the feature representation to the two target classes (crack and background). All convolutions in the model are implemented without bias terms, as each is directly followed by a Batch Normalization layer.

\subsection{Dataset and Preprocessing}
In this work, we use the CrackVision12K dataset \cite{goo2025hybrid}. CrackVision12K is a curated benchmark for crack segmentation that combines images from several previously published crack datasets into a unified dataset with improved and standardized annotations. Compared with the original source datasets, the labels were manually refined to reduce annotation noise and inconsistencies. In addition, all images and masks were resized to a common spatial resolution of $256 \times 256$ pixels.

In total, the dataset contains 12000 image-mask pairs. To ensure comparability with prior work on CrackVision12K, we use the train/validation/test split provided in \cite{goo2025hybrid}. The training set contains 9600 images, while the validation and test sets each contain 1200 images.

\subsection{Data Augmentation}
We also applied data augmentation methods during training to further improve the segmentation results and make the models more robust for practical applications. Each augmentation is applied with a probability of 50\%. The augmentation pipeline includes additive image noise, motion blur, horizontal flipping, and small random rotations. 

\subsection{Knowledge Distillation} \label{sec:knowledge_distillation}
To improve the segmentation results even more, we use \gls{kd}, where a stronger teacher network guides the training of a smaller student network \cite{hinton2015distilling}. In our setup, the teacher is a SegFormer-B5 model \cite{xie2021segformer} that is first fine-tuned on CrackVision12K. The student is one of our channel-scaled U-Net variants. During KD training, the teacher remains frozen, while the student is optimized.

Let $x \in \mathbb{R}^{H \times W \times 3}$ denote an input image and $y$ its ground-truth segmentation mask. The teacher produces logits $z_t(x)$ and the student produces logits $z_s(x)$. Since the spatial resolution of the teacher and student models is not equal, we resize the teacher logits to match the student's output resolution by using bilinear interpolation. Afterwards, we apply KD on the softened class distributions:
\begin{equation}
p_t(x) = \mathrm{softmax}\left(\tfrac{z_t(x)}{T}\right), \hspace*{2mm} p_s(x) = \mathrm{softmax}\left(\tfrac{z_s(x)}{T}\right)
\end{equation}
where $T > 1$ is the temperature parameter.

The KD loss is defined as the Kullback-Leibler divergence between teacher and student predictions:
\begin{equation}
\mathcal{L}_{KD} = T^2 \cdot \mathrm{KL}\!\left(p_t(x)\,\|\,p_s(x)\right).
\end{equation}

In addition to the soft teacher supervision, we use a hard loss based on the ground-truth labels. More specifically, we combine weighted cross-entropy with a soft Dice loss computed on the crack class:
\begin{equation}
\mathcal{L}_{hard} = \mathcal{L}_{CE} + \lambda_{Dice}\,\mathcal{L}_{Dice}
\end{equation}
The factor $\lambda_{Dice}$ weights the impact of the Dice loss and is determined by our hyperparameter search. Thus, our final training loss is then:
\begin{equation}
\mathcal{L} = \alpha \mathcal{L}_{KD} + (1-\alpha) \mathcal{L}_{hard},
\end{equation}
where $\alpha \in [0,1]$ controls the trade-off between teacher supervision and supervision from the ground-truth labels.

\subsection{Training Setup and Hyperparameter Search}

We first train all five U-Net variants with standard supervised learning to obtain the baselines. Then we optimize Hyperparameters with a combination of random search and an evolutionary algorithm, adopted from \cite{tschope2025novel}. The search covers the AdamW \cite{loshchilov2017decoupled} learning rate, momentum parameters, and weight decay, the scheduler decay $\gamma$ and step size, and the training-time class weights. For KD training, we additionally optimize the temperature $T$, the mixing factor $\alpha$, and the Dice weight $\lambda_{Dice}$. Training-time class weights are not used at evaluation. Instead, we use the fixed weights described in \Cref{sec:evaluation_metrics} for all reported numbers.

For training with \gls{kd}, we additionally optimize the \gls{kd} temperature $T$, the KD mixing factor $\alpha$, and the weighting factor of the Dice term $\lambda_{Dice}$. The hard supervision consists of weighted cross-entropy and a soft Dice loss on the crack class, while the soft supervision is given by the KL divergence between teacher and student predictions.

All models are trained for at most 150 epochs with early stopping. For each model configuration, we repeat the training with five random seeds and report the mean and standard deviations of the evaluation metrics.

\subsection{Evaluation Metrics} \label{sec:evaluation_metrics}
Crack segmentation is in general a task with a highly imbalanced class distribution, as cracks are often very thin and small and occur much less frequently than the background. This is also the case with the CrackVision12K dataset that we use. If we only used the mean \gls{iou} as a metric, we would weight the background class and the much more interesting crack class equally, which would end in a deceptively high evaluation score. Therefore, in addition to mean \gls{iou}, we also use weighted \gls{iou}. Below, we present the definitions of both metrics in more detail.

Let $\mathrm{IoU}_{bg}$ and $\mathrm{IoU}_{crack}$ denote the \gls{iou} values for the background and crack classes, respectively. The mean \gls{iou} is then defined as:
\begin{equation}
\mathrm{mIoU} = 0.5 \cdot \mathrm{IoU}_{bg} + 0.5 \cdot \mathrm{IoU}_{crack}
\end{equation}
The weighted \gls{iou} instead emphasizes the crack class more strongly than the background class:
\begin{equation}
\mathrm{wIoU} = w_{bg} \cdot \mathrm{IoU}_{bg} + w_{crack} \cdot \mathrm{IoU}_{crack}
\end{equation}
with
\begin{equation}
w_{bg}=0.068,\qquad w_{crack}=0.932.
\end{equation}
We compute the weights from the inverse class frequencies of the training set and normalize them so that they sum to one. The weights are fixed for all our experiments and are therefore not part of the hyperparameter search. A large gap between mean \gls{iou} and weighted \gls{iou} of a model means that most of its mean \gls{iou} comes from the background pixels and not from the crack pixels themselves.

\section{Implementation}

In this paper we investigate the performance of U-Net implementations on four representative edge computing platforms: a conventional CPU, an edge GPU, a dedicated AI accelerator (TPU), and a custom FPGA design. The selected platforms cover a wide range of edge hardware under different computational and energy constraints.

\subsection{Edge Platforms}

We consider three edge platforms: The Raspberry Pi 5 Model B (8 GB) with a quad-core Cortex-A76 CPU, the Seeed Studio reComputer J3011 with an NVIDIA Jetson Orin Nano (8 GB), and the Coral Edge TPU M.2 module hosted on a Raspberry Pi 5. On the CPU we run the U-Net via ONNX Runtime in fp32. On the GPU we use TensorRT in two configurations: full fp32, and int8-weights with fp32-activations. On the Edge TPU we deploy fully int8-quantized TensorFlow Lite models compiled with the Edge TPU compiler.

To measure inference latency, we use a batch size of 1 and calculate the average across five execution runs on a subset of 1,024 input images, including host–device transfer and fp32/int8 conversion overhead. Power is measured with an external USB POWER-Z KM003C KT002 multimeter, both at idle and during 20 s of continuous inference, averaged over three measurements. For the Coral Edge TPU, the measurement covers both the M.2 module and its Raspberry Pi 5 host. All Raspberry Pi devices use active cooling; the Edge TPU uses an aluminum heat sink.

\subsection{Custom Hardware Implementation on FPGA}

Our \gls{fpga} implementation uses an \gls{hls} hardware library that provides custom hardware architectures for many common \gls{dnn} layers and components. The library is built as a set of C++ template functions with \gls{hls} annotations, which makes it easy to support different network topologies and to explore the design space quickly. We design the architecture for low power and low latency. To this end, we keep all weights and most intermediate results in on-chip memory, since off-chip transfers cost more energy and add latency.

\begin{figure}[!h]
\centering
\includegraphics[width=\columnwidth]{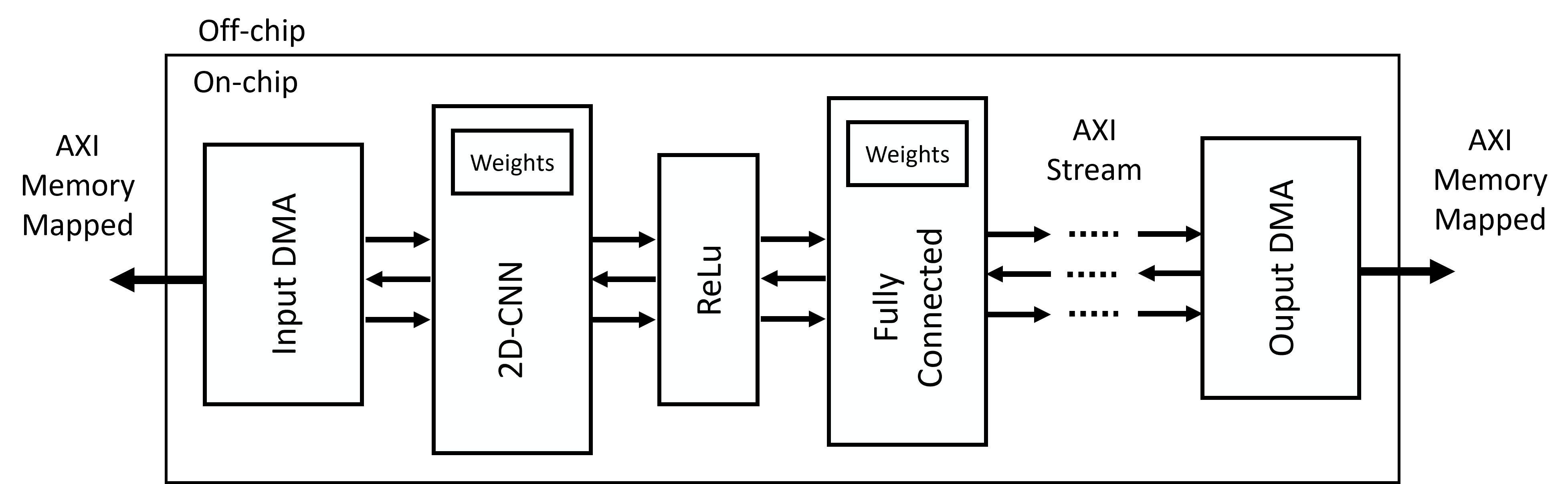}
\caption{The sample hardware architecture where each layer is mapped to a separate hardware instance. The modules are connected via on-chip data streams.}
\label{fig:hw_architecture}
\end{figure}

Nevertheless, a design relying solely on on-chip resources inherently limits the size of \gls{dnn} models that can be feasibly implemented. To overcome this constraint, we developed a hybrid hardware architecture that supports off-chip implementation of skip and residual connections, effectively relaxing the limitations on model size and structural complexity. The architecture is fully pipelined, allowing all layers to operate concurrently and initiate computation immediately once their inputs become available, thereby reducing both latency and energy consumption. Individual hardware modules corresponding to each layer are interconnected via on-chip data streams and assembled into a single top-level module, as illustrated in \cref{fig:hw_architecture}. 

Power was measured using the onboard Texas Instruments IN219 current/power monitor. Idle power was measured after configuring the PL for a continuous period of 10 seconds. While the runtime power was averaged over 32 repetitions, each consisting of 32 images processed in series. The measurement includes the PL reading the image from the memory, processing it, and writing the result back to the memory. The same methodology was used for measuring latency.

\section{Results}

\subsection{Scope and Fairness of the Cross-Platform Comparison} 
The comparison in \Cref{tab:edge_computing_platforms} is a practical evaluation rather than a strictly controlled evaluation. The used models differ across platforms in three ways, each of which is due more to platform-specific constraints rather than to the design itself. First, models targeted at the Coral Edge TPU replace transposed convolutions with nearest-neighbor upsampling to avoid the accuracy drop we observed on this device; All other platforms retain the original transposed-convolution decoder. Second, the numerical precision is not uniform: the RPi5 runs fp32 weights and activations, the Jetson Orin Nano is evaluated in both fp32 and a mixed int8-weights/fp32-activations configuration (native to TensorRT), the Edge TPU and the FPGA executes fully int8. Third, the measurement instrumentation differs between platforms. We therefore interpret the numbers as realistic deployment results for each platform in its best-supported configuration, rather than as an apples-to-apples comparison of the same model across devices. \label{sec:comparison_scope}

\subsection{Accuracy-Efficiency Trade-off Analysis}
\Cref{tab:results_general} summarizes the segmentation performance of all models tested in the CrackVision12K dataset before quantization. All models shown in \cref{tab:results_general} were trained on five different random seeds. Therefore, the mean and standard deviation for the weighted \gls{iou} and the unweighted \gls{iou} are provided.

Considering the model results without knowledge distillation, the original U-Net ($c=32$) serves as the baseline. Reducing the base channel count to $c=16$ reduces the number of \gls{mac} operations by approximately $4\times$, from 24.21G to 6.10G, while the accuracy drop is negligible - only 0.18\glspl{pp} in weighted \gls{iou} and 0.08\glspl{pp} in mean \gls{iou}. Reducing further to $c=8$ provides an additional $\sim4\times$ MAC reduction to approximately 1.55G, with a still modest accuracy cost of only 3.63\glspl{pp} in weighted \gls{iou} and 2.11\glspl{pp} in mean \gls{iou} relative to the baseline - a competitive result given the overall $16\times$ reduction in computational cost. Scaling down to $c=2$, however, incurs steep accuracy costs: a drop of 13.60\glspl{pp} in weighted \gls{iou} and 10.46\glspl{pp} in mean \gls{iou} compared to $c=16$, confirming that $c=16$ and $c=8$ represent the best efficiency--accuracy trade-offs in this family. As expected, SegFormer B5 outperforms all U-Net variants without knowledge distillation, owing to its considerably larger capacity and modern transformer-based architecture. 

With knowledge distillation, all U-Net variants improve. The baseline model ($c=32$) achieves a weighted \gls{iou} of $51.06\%$ and a mean \gls{iou} of $71.92\%$---within 2.07\glspl{pp} of the teacher SegFormer B5 ($53.13\%$ weighted \gls{iou}), showing that KD brings the student nearly to the teacher's level. Crucially, the efficiency gains identified without KD are preserved: $c=16$ achieves $50.63\%$ weighted \gls{iou} and $71.74\%$ mean \gls{iou}, only 0.43 and 0.18\glspl{pp} below $c=32$, while still benefiting from the same $4\times$ MAC reduction. Similarly, $c=8$ achieves $49.25\%$ weighted \gls{iou} and $70.96\%$ mean \gls{iou}---only 1.81\glspl{pp} and 0.96\glspl{pp} below the best U-Net model ---at a $16\times$ overall reduction in compute.

Overall, these results show that KD not only improves segmentation results across all U-Net variants, but also makes the models significantly more robust to channel reduction: the mean \gls{iou} gap between $c=32$ and $c=8$ shrinks from 2.19\glspl{pp} without KD to only 0.96\glspl{pp} with KD, which shows that KD compensates for the capacity loss introduced by pruning. In the next subsection, we evaluate each variant across several edge platforms under quantization.

\setlength{\tabcolsep}{4pt}

\begin{table}[!t]
\begin{threeparttable}
\centering
\caption{Comparison of U-Net variants with different channel scaling factors, with and without knowledge distillation (KD). The Hybrid-Segmentor model was proposed in \cite{goo2025hybrid}. SegFormer B5 is included as a high-capacity reference.}
\label{tab:results_general}

{%
\begin{tabular}{l|cccc}
\toprule
Approach & \begin{tabular}[c]{@{}c@{}}Weighted\\ \gls{iou} {[}\%{]}\end{tabular} & \begin{tabular}[c]{@{}c@{}}Mean\\ \gls{iou} {[}\%{]}\end{tabular} & \begin{tabular}[c]{@{}c@{}}GOPs \end{tabular} & \begin{tabular}[c]{@{}c@{}}Num. \\ Params (M)\end{tabular} \\
\cmidrule{1-5}
Hybrid-Segmentor & - & $63.1 \pm0.36 $  & - & 237.86 \\
\cmidrule{1-5}
SegFormer B5 & $53.13 \pm0.21$ & $73.42 \pm0.26$  & $51.25$ & $84.59$ \\
\cmidrule{1-5}
Base 2 w/o KD  & $29.31 \pm0.54$ & $56.17 \pm0.43$ & $0.11$ & $0.031$ \\
Base 4 w/o KD  & $37.67 \pm0.43$ & $62.54 \pm0.32$ & $0.40$ & $0.122$ \\
Base 8 w/o KD  & $39.10 \pm1.13$ & $64.44 \pm0.81$ & $1.55$ & $0.487$ \\
Base 16 w/o KD & $42.73 \pm1.50$ & $66.55 \pm1.06$ & $6.10$ & $1.943$ \\
Base 32 w/o KD\,* & $42.91 \pm0.56$ & $66.63 \pm0.43$ & $24.21$ & $7.763$ \\
\cmidrule{1-5}
Base 2 KD  & $47.54 \pm0.31$ & $69.90 \pm0.27$ & $0.11$ & $0.031$ \\
Base 4 KD  & $48.81 \pm0.95$ & $70.68 \pm0.78$ & $0.40$ & $0.122$ \\
Base 8 KD  & $49.25 \pm0.67$ & $70.96 \pm0.59$ & $1.55$ & $0.487$ \\
Base 16 KD & $50.63 \pm0.55$ & $71.74 \pm0.44$ & $6.10$ & $1.943$ \\
Base 32 KD & $\mathbf{51.06 \pm0.71}$ & $\mathbf{71.92 \pm0.65}$ & $\mathbf{24.21}$ & $\mathbf{7.763}$ \\
\bottomrule
\end{tabular}
}

\begin{tablenotes}
\footnotesize
\item * This model is used as a baseline.
\end{tablenotes}
\end{threeparttable}
\end{table}

\subsection{Hardware Implementation Results}
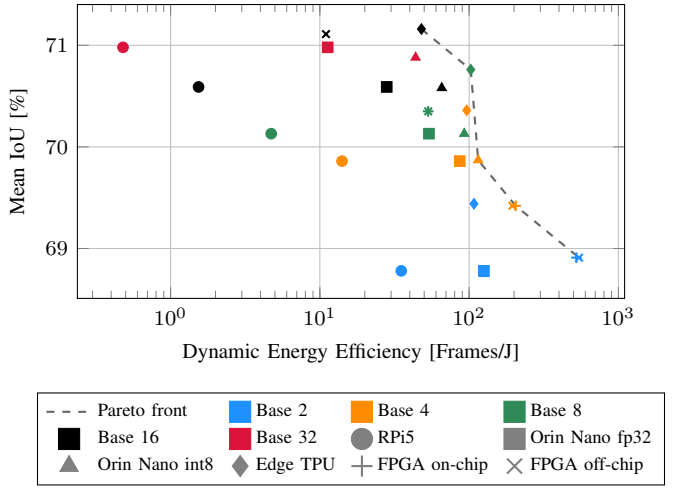
\begin{figure}[!h]
\centering

\begin{tikzpicture}

\definecolor{dodgerblue}{RGB}{30, 144, 255}
\definecolor{darkorange}{RGB}{255, 140, 0}
\definecolor{seagreen}{RGB}{46, 139, 87}
\definecolor{crimson}{RGB}{220, 20, 60}

\begin{axis}[
    width=\columnwidth,
    height=0.62\columnwidth,
    xmode=log,
    xlabel={Dynamic Energy Efficiency [Frames/J]},
    ylabel={Mean IoU [\%]},
    label style={font=\footnotesize},
    tick label style={font=\footnotesize},
    mark size=2pt,
    grid=major,
    legend columns=4,
    legend style={at={(0.5,-0.32)}, anchor=north, font=\scriptsize,
                  /tikz/every even column/.append style={column sep=4pt}},
    legend cell align={left},
]
\addplot[only marks, mark=*, color=dodgerblue, fill=dodgerblue, forget plot]
    coordinates {(35.09, 68.78)};
\addplot[only marks, mark=square*, color=dodgerblue, fill=dodgerblue, forget plot]
    coordinates {(125.58, 68.78)};
\addplot[only marks, mark=triangle*, color=dodgerblue, fill=dodgerblue, forget plot]
    coordinates {(126.4, 68.75)};
\addplot[only marks, mark=diamond*, color=dodgerblue, fill=dodgerblue, forget plot]
    coordinates {(107.77, 69.44)};
\addplot[only marks, mark=+, color=dodgerblue, thick, forget plot]
    coordinates {(522.98, 68.91)};
\addplot[only marks, mark=x, color=dodgerblue, thick, forget plot]
    coordinates {(544.65, 68.91)};

\addplot[only marks, mark=*, color=darkorange, fill=darkorange, forget plot]
    coordinates {(14.08, 69.86)};
\addplot[only marks, mark=square*, color=darkorange, fill=darkorange, forget plot]
    coordinates {(86.58, 69.86)};
\addplot[only marks, mark=triangle*, color=darkorange, fill=darkorange, forget plot]
    coordinates {(114.97, 69.87)};
\addplot[only marks, mark=diamond*, color=darkorange, fill=darkorange, forget plot]
    coordinates {(96.35, 70.36)};
\addplot[only marks, mark=+, color=darkorange, thick, forget plot]
    coordinates {(204.99, 69.42)};
\addplot[only marks, mark=x, color=darkorange, thick, forget plot]
    coordinates {(194.90, 69.42)};

\addplot[only marks, mark=*, color=seagreen, fill=seagreen, forget plot]
    coordinates {(4.72, 70.13)};
\addplot[only marks, mark=square*, color=seagreen, fill=seagreen, forget plot]
    coordinates {(53.81, 70.13)};
\addplot[only marks, mark=triangle*, color=seagreen, fill=seagreen, forget plot]
    coordinates {(92.75, 70.13)};
\addplot[only marks, mark=diamond*, color=seagreen, fill=seagreen, forget plot]
    coordinates {(102.56, 70.76)};
\addplot[only marks, mark=+, color=seagreen, thick, forget plot]
    coordinates {(53.16, 70.35)};
\addplot[only marks, mark=x, color=seagreen, thick, forget plot]
    coordinates {(53.23, 70.35)};

\addplot[only marks, mark=*, color=black, fill=black, forget plot]
    coordinates {(1.54, 70.59)};
\addplot[only marks, mark=square*, color=black, fill=black, forget plot]
    coordinates {(28.06, 70.59)};
\addplot[only marks, mark=triangle*, color=black, fill=black, forget plot]
    coordinates {(65.56, 70.58)};
\addplot[only marks, mark=diamond*, color=black, fill=black, forget plot]
    coordinates {(47.92, 71.16)};
\addplot[only marks, mark=x, color=black, thick, forget plot]
    coordinates {(10.99, 71.11)};

\addplot[only marks, mark=*, color=crimson, fill=crimson, forget plot]
    coordinates {(0.48, 70.98)};
\addplot[only marks, mark=square*, color=crimson, fill=crimson, forget plot]
    coordinates {(11.29, 70.98)};
\addplot[only marks, mark=triangle*, color=crimson, fill=crimson, forget plot]
    coordinates {(43.79, 70.88)};

\addplot[dashed, thick, color=black!60, forget plot]
    coordinates {
        (47.92, 71.16)
        (102.56, 70.76)
        (114.97, 69.87)
        (204.99, 69.42)
        (544.65, 68.91)
    };

\addlegendimage{dashed, thick, color=black!60}
\addlegendentry{Pareto front}
\addlegendimage{only marks, mark=square*, mark size=4pt, color=dodgerblue, fill=dodgerblue}
\addlegendentry{Base 2}
\addlegendimage{only marks, mark=square*, mark size=4pt, color=darkorange, fill=darkorange}
\addlegendentry{Base 4}
\addlegendimage{only marks, mark=square*, mark size=4pt, color=seagreen, fill=seagreen}
\addlegendentry{Base 8}
\addlegendimage{only marks, mark=square*, mark size=4pt, color=black, fill=black}
\addlegendentry{Base 16}
\addlegendimage{only marks, mark=square*, mark size=4pt, color=crimson, fill=crimson}
\addlegendentry{Base 32}
\addlegendimage{only marks, mark=*, mark size=4pt, color=gray, fill=gray}
\addlegendentry{RPi5}
\addlegendimage{only marks, mark=square*, mark size=4pt, color=gray, fill=gray}
\addlegendentry{Orin Nano fp32}
\addlegendimage{only marks, mark=triangle*, mark size=4pt, color=gray, fill=gray}
\addlegendentry{Orin Nano int8}
\addlegendimage{only marks, mark=diamond*, mark size=4pt, color=gray, fill=gray}
\addlegendentry{Edge TPU}
\addlegendimage{only marks, mark=+, mark size=4pt, color=gray, thick}
\addlegendentry{FPGA on-chip}
\addlegendimage{only marks, mark=x, mark size=4pt, color=gray, thick}
\addlegendentry{FPGA off-chip}

\end{axis}
\end{tikzpicture}
\caption{Pareto frontier showing the trade-off between accuracy (weighted IoU) and energy efficiency across different platforms and model sizes. FPGA on-chip: U-Net skip connections stored on chip, FPGA off-chip: U-Net skip connections stored off chip in the memory}
\label{fig:pareto_frontier}
\end{figure}

As discussed in \Cref{sec:comparison_scope}, the configurations per platform are not identical, the comparison reflects the best-supported deployment path on each device. \Cref{tab:edge_computing_platforms} shows the results for knowledge-distilled models across three edge platforms and the FPGA, reporting segmentation accuracy (weighted and mean \gls{iou}), throughput (\gls{fps}), and power and energy metrics. In addition to idle and runtime power, both dynamic and runtime energy efficiency (Frames/J) are provided. Our analysis focuses primarily on dynamic energy efficiency, as it isolates the energy consumption attributable to active inference. In contrast, runtime energy efficiency is affected by background processes and system-level overhead, making cross-platform comparisons less reliable. Runtime energy efficiency does not solve all measurement issues, but it gives a more realistic picture of how the models behave in deployment. \Cref{fig:pareto_frontier} further illustrates the trade-off between dynamic energy efficiency and mean IoU across the evaluated configurations.

\subsubsection{FPGA results}
For an efficient \gls{fpga} implementation \gls{ptq} is used to reduce the precision of the model parameters. Two experiments were conducted; in the first experiment, the weights and activations were quantized to int8 as shown in \Cref{tab:quantization_results} this experiment achieved a similar or higher \gls{iou} compared to the fp32 experiment due to regularization. In the second experiment, the weights were quantized to int4, which resulted in a considerable drop in accuracy as shown in \Cref{tab:quantization_results}. We see this drop as a limit of \gls{ptq} at very low bit widths and leave a follow-up with \gls{qat} to future work.
Based on the results in \Cref{tab:quantization_results} int8 models were chosen for implementation on \gls{fpga}, for which the AMD Vitis 2024.2 \gls{hls} tool was used and the target platform was set to TySOM-2A-ZU19EG. \Cref{tab:fpga_resource_utilization} presents the resources required to implement the chosen models on the target platform. Two experiments were conducted; in the first experiment, the resource-heavy skip connections of the U-Net model were kept on chip, while in the second experiment, they were moved to off-chip memory. This made it possible to implement the larger model with $c=16$. Skip connections were moved to off-chip memory by creating an AXI interface for each skip connection, which is used for writing and reading. This is required due to the high  write and read speeds and the fact that all skip connections are accessed in parallel. On-chip buffers were added to the design to bridge the gap between the high-speed \gls{pl} and the slow off-chip memory. The extra buffers and interfaces lead to an average increase of 2.19\glspl{pp} and 1.60\glspl{pp} in \gls{lut} and \gls{ff} respectively, which is within the acceptable overhead. This shows the tradeoff between moving skip connections off chip and keeping them on chip, which is less efficient for small models i.e., with $c=2$. \Cref{tab:edge_computing_platforms} shows the throughput, \gls{iou}, energy and efficiency of the implemented models on the \gls{fpga}. All \gls{fpga} implementations achieved the same \gls{iou} as the corresponding software implementations.

\subsubsection{Edge Computing Platforms}

The results in \Cref{tab:edge_computing_platforms} show that the models on the edge platforms achieve \gls{iou} values close to the corresponding software results from \Cref{tab:results_general}. This again shows that the reduction of the model size from Base~32 to Base~2 - which corresponds to a 256$\times$ reduction in the number of parameters - leads to less than 2\glspl{pp} loss in the mean IoU on the edge platforms. The Raspberry Pi 5 reflects the limitations of CPU-based inference, since it no longer achieves real-time performance for the larger models. In contrast, the Jetson Orin Nano reaches real-time performance for all tested model sizes. The comparison between the fp32 and int8 runs on the Orin Nano also shows the benefit of quantization, especially in terms of dynamic energy efficiency.

Although the Coral Edge TPU supports transposed convolutions, models that rely on transposed convolution for upsampling experience a significant drop in accuracy when deployed on the Coral Edge TPU. To address this, the models were retrained using nearest upsampling, which preserved their accuracy. However, the model with base 32 failed to run on the Coral Edge TPU.

\setlength{\tabcolsep}{4pt}
\begin{table}[!t]
  \centering
  \caption{Post-training quantization results for models with knowledge distillation.} 
    \label{tab:quantization_results}
\begin{tabular}{c|ccccc}
\toprule
\begin{tabular}[c]{@{}c@{}}Weight\\ Precision\end{tabular} & Base & \begin{tabular}[c]{@{}c@{}}Activation\\ Precision\end{tabular} & \begin{tabular}[c]{@{}c@{}}Weighted\\ \gls{iou} {[}\%{]}\end{tabular} & \begin{tabular}[c]{@{}c@{}}Mean\\ \gls{iou} {[}\%{]}\end{tabular} \\
\midrule
\multirow{5}{*}{int8}
 & 2  & 8 & 45.49 & 68.91 \\
 & 4  & 8 & 46.29 & 69.42 \\
 & 8  & 8 & 47.86 & 70.35 \\
 & 16 & 8 & 49.22 & 71.11 \\
 & 32 & 8 & 49.05 & 70.92 \\
\specialrule{1pt}{1pt}{0.4pt}
\specialrule{0.4pt}{0.4pt}{2pt}
\multirow{5}{*}{int4}
 & 2  & 8 & 40.62 & 65.09 \\
 & 4  & 8 & 42.37 & 66.76 \\
 & 8  & 8 & 42.87 & 67.32 \\
 & 16 & 8 & 38.23 & 64.85 \\
 & 32 & 8 & 48.46 & 70.73 \\
\bottomrule
\end{tabular}
\end{table}

\setlength{\tabcolsep}{4pt}
\begin{table}[!t]
\centering
\caption{Implementation Resource Utilization (\%) on AMD Zynq UltraScale+ ZU19EG-FFVB1517 FPGA.}
\label{tab:fpga_resource_utilization}
\begin{tabular}{c|ccccc}
\toprule
Skip & Base & \begin{tabular}[c]{@{}c@{}}LUTs\\ {[}\%{]}\end{tabular} & \begin{tabular}[c]{@{}c@{}}FFs\\ {[}\%{]}\end{tabular} & \begin{tabular}[c]{@{}c@{}}BRAMs 36K\\ {[}\%{]}\end{tabular} & \begin{tabular}[c]{@{}c@{}}DSPs\\ {[}\%{]}\end{tabular} \\
\midrule
\multirow{3}{*}{On-chip} 
 & 2  & 11.65 & 3.35 & 15.24 & 11.18 \\
 & 4  & 16.81 & 4.51 & 33.23 & 43.14 \\
 & 8  & 23.45 & 5.88 & 78.86 & 92.17 \\
\specialrule{1pt}{1pt}{0.4pt}
\specialrule{0.4pt}{0.4pt}{2pt}
\multirow{4}{*}{Off-chip}
 & 2  & 13.90 & 4.94 & 12.40  & 11.18  \\
 & 4  & 19.13 & 6.14 & 28.00  & 43.14 \\
 & 8  & 25.48 & 7.47 & 55.79  & 92.17 \\
 & 16 & 36.88 & 8.07 & 100.00 & 97.00 \\
\bottomrule
\end{tabular}
\end{table}

\begin{table*}[!t]
\centering
\caption{Comparison of throughput ($\theta$), \gls{iou}, and energy efficiency for various model precisions on RPi5, Orin Nano, Edge TPU, and FPGA with skip connections mapped to on-chip and off-chip memory at batch size 1.}
\label{tab:edge_computing_platforms}
\begin{tabular}{c|cccccccccc}
\toprule
Base & Device & \begin{tabular}[c]{@{}c@{}}Model\\{[}bits{]}\end{tabular} & \begin{tabular}[c]{@{}c@{}}Data\\{[}bits{]}\end{tabular} & \begin{tabular}[c]{@{}c@{}}Weighted\\ \gls{iou}\\ {[}\%{]}
\end{tabular} & \begin{tabular}[c]{@{}c@{}}Mean\\ \gls{iou}\\ {[}\%{]}\end{tabular} & \begin{tabular}[c]{@{}c@{}}$\theta$\\ {[}\gls{fps}{]}\end{tabular} & \begin{tabular}[c]{@{}c@{}}Idle\\ Power\\ {[}W{]}\end{tabular} & \begin{tabular}[c]{@{}c@{}}Runtime\\ Power\\ {[}W{]}\end{tabular} & \begin{tabular}[c]{@{}c@{}}Dynamic\\ Energy\\ Efficiency\\ {[}Frames/J{]}\end{tabular} & \begin{tabular}[c]{@{}c@{}}Runtime\\ Energy\\ Efficiency\\ {[}Frames/J{]}\end{tabular} \\
\midrule

\multirow{6}{*}{2} 
 & RPi5 & fp32 & fp32 & 45.28 & 68.78 & 246 & 2.73 & 9.74 & 35.09 & 25.26 \\
\cmidrule{2-11}
 & \multirow{3}{*}{Orin Nano} & fp32 & fp32 & 45.28 & 68.78 & 378 & 6.97 & 9.98 & 125.58 & 37.88 \\
 \cmidrule{3-11}
 & & int8 & fp32 & 45.18 & 68.75 & 383 & 6.97 & 10.00 & 126.40 & 38.30 \\
 \cmidrule{2-11}
 & Edge TPU & int8 & int8 & 46.31 & 69.44 & 222 & 3.38 & 5.44 & 107.77 & 40.81 \\
 \cmidrule{2-11}
 & FPGA (on-chip) & int8 & int8 & 45.49 & 68.91 & 397 & 18.71 & 19.47 & 522.98 & 20.41 \\
 \cmidrule{2-11}
 & FPGA (off-chip) & int8 & int8 & 45.49 & 68.91 & \textbf{398} & 18.74 & 19.47 & \textbf{544.65} & 20.42 \\
\specialrule{1pt}{1pt}{0.4pt}  
\specialrule{0.4pt}{0.4pt}{2pt} 

\multirow{6}{*}{4}
 & RPi5 & fp32 & fp32 & 47.32 & 69.86 & 98 & 2.73 & 9.69 & 14.08 & 10.11 \\
\cmidrule{2-11}
 & \multirow{3}{*}{Orin Nano} & fp32 & fp32 & 47.32 & 69.86 & 329 & 6.97 & 10.77 & 86.58 & 30.55 \\
  \cmidrule{3-11}
 & & int8 & fp32 & 47.34 & 69.87 & 361 & 6.97 & 10.11 & 114.97 & 35.71 \\
  \cmidrule{2-11}
 & Edge TPU & int8 & int8 & 47.95 & 70.36 & 211 & 3.27 & 5.46 & 96.35 & 38.64 \\
  \cmidrule{2-11}
 & FPGA (on-chip) & int8 & int8 & 46.29 & 69.42 & \textbf{398} & 18.89 & 20.83 & \textbf{204.99} & 19.09 \\
  \cmidrule{2-11}
 & FPGA (off-chip) & int8 & int8 & 46.29 & 69.42 & 398 & 19.08 & 21.12 & 194.90 & 18.83 \\
\specialrule{1pt}{1pt}{0.4pt}  
\specialrule{0.4pt}{0.4pt}{2pt} 

\multirow{6}{*}{8}
 & RPi5 & fp32 & fp32 & 47.56 & 70.13 & 34 & 2.73 & 9.93 & 4.72 & 3.42 \\
\cmidrule{2-11}
 & \multirow{3}{*}{Orin Nano} & fp32 & fp32 & 47.56 & 70.13 & 247 & 6.97 & 11.56 & 53.81 & 21.37 \\
  \cmidrule{3-11}
 & & int8 & fp32 & 47.56 & 70.13 & 320 & 6.97 & 10.42 & 92.75 & 30.71 \\
 \cmidrule{2-11}
 & Edge TPU & int8 & int8 & 48.59 & 70.76 & 160 & 3.27 & 4.83 & \textbf{102.56} & 33.13 \\
 \cmidrule{2-11}
 & FPGA (on-chip) & int8 & int8 & 47.86 & 70.35 & 397 & 19.29 & 26.76 & 53.16 & 14.84 \\
 \cmidrule{2-11}
 & FPGA (off-chip) & int8 & int8 & 47.86 & 70.35 & \textbf{398} & 19.32 & 26.79 & 53.23 & 14.84 \\
\specialrule{1pt}{1pt}{0.4pt}  
\specialrule{0.4pt}{0.4pt}{2pt} 

\multirow{5}{*}{16}
 & RPi5 & fp32 & fp32 & 48.50 & 70.59 & 12 & 2.73 & 10.54 & 1.54 & 1.14 \\
\cmidrule{2-11}
 & \multirow{3}{*}{Orin Nano} & fp32 & fp32 & 48.50 & 70.59 & 158 & 6.97 & 12.60 & 28.06 & 12.54 \\
  \cmidrule{3-11}
 & & int8 & fp32 & 48.50 & 70.58 & \textbf{276} & 6.97 & 11.18 & \textbf{65.56} & 24.69 \\
 \cmidrule{2-11}
 & Edge TPU & int8 & int8 & 49.26 & 71.16 & 69 & 3.27 & 4.71 & 47.92 & 14.65 \\
 \cmidrule{2-11}
 & FPGA (off-chip) & int8 & int8 & 49.22 & 71.11 & 82 & 19.75 & 27.22 & 10.99 & 3.02 \\
\specialrule{1pt}{1pt}{0.4pt}  
\specialrule{0.4pt}{0.4pt}{2pt} 

\multirow{4}{*}{32}
 & RPi5 & fp32 & fp32 & 49.22 & 70.98 & 4 & 2.73 & 11.06 & 0.48 & 0.36 \\
\cmidrule{2-11}
 & \multirow{3}{*}{Orin Nano} & fp32 & fp32 & 49.22 & 70.98 & 78 & 6.97 & 13.88 & 11.29 & 5.62 \\
  \cmidrule{3-11}
 & & int8 & fp32 & 49.06 & 70.88 & \textbf{222} & 6.97 & 12.04 & \textbf{43.79} & 18.44 \\
\bottomrule
\end{tabular}
\end{table*}

\section{Conclusion and Future Work}

In this work, we studied crack segmentation on resource-constrained edge platforms from two sides: The model design and the target hardware. For this, we scaled the U-Net architecture over three orders of magnitude in parameter count and analyzed the trade-off between \gls{iou} and computational cost. Our results show that reducing the model size can strongly lower the computational cost, while some of the smaller variants still remain close to the baseline in terms of segmentation accuracy. To improve our scaled U-Net models further, we applied knowledge distillation by using SegFormer B5 as the teacher model. This improved all U-Net variants.

We then evaluated all variants on several edge platforms and compared their throughput, power consumption, and energy efficiency. We also developed a custom \gls{fpga}-based architecture for single-batch inference. Among all tested platforms, the FPGA (on-chip) implementation with the U-Net Base 4 using \gls{kd} and int8 quantization achieves an energy efficiency of 204.99 Frames/J and a throughput of 398 \gls{fps}, while improving the mean \gls{iou} by 2.79\glspl{pp} compared to the baseline model (U-Net Base 32 without \gls{kd}). Compared to the Hybrid-Segmentor model from \cite{goo2025hybrid}, which is - to the best of our knowledge - currently the best tested model on this dataset, our aforementioned U-Net Base 4 model with \gls{kd} is 6.32\glspl{pp} higher in mean \gls{iou} by using the same train/val/test split.

Therefore, our results show that the combination of model scaling, knowledge distillation, quantization, and hardware-specific implementation is useful for crack segmentation on edge devices. 

For future work, we plan to investigate quantization-aware training and structured pruning to reduce the model cost further. In addition, it would be interesting to study other teacher models or multi-teacher settings for knowledge distillation. Finally, we plan to record and annotate our own crack segmentation dataset from \gls{uav} flights to evaluate the models under more realistic conditions.

\balance
\bibliographystyle{IEEEtran}
\bibliography{references}
\end{document}